\documentclass[graybox]{svmult}

\usepackage{mathptmx}       
\usepackage{helvet}         
\usepackage{courier}        
\usepackage{type1cm}        
\usepackage{makeidx}         
\usepackage{graphicx}        
\usepackage{multicol}        
\usepackage[bottom]{footmisc}
\usepackage{natbib}

\makeindex             

\begin{document}

\title*{Closing in on the Cosmos: Cosmology's Rebirth and the Rise of the Dark Matter Problem 
}
\titlerunning{Closing in on the Cosmos} 
\author{Jaco de Swart}
\institute{Jaco de Swart \at Institute of Physics \& Vossius Center for the History of Humanities and Sciences \\ University of Amsterdam, Science Park 904, Amsterdam, The Netherlands \\ \email{j.g.deswart@uva.nl}
}

\maketitle

\abstract{Influenced by the renaissance of general relativity that came to pass in the 1950s, the character of cosmology fundamentally changed in the 1960s as it became a well-established empirical science. Although observations went to dominate its practice, extra-theoretical beliefs and principles reminiscent of methodological debates in the 1950s kept playing an important tacit role in cosmological considerations. Specifically, belief in cosmologies that modeled a ``closed universe'' based on Machian insights remained influential. The rise of the dark matter problem in the early 1970s serves to illustrate this hybrid methodological character of cosmological science. 
}


\section*{Introduction}

In 1974, two landmark papers were published by independent research groups in the U.S. and Estonia that concluded on the existence of \textit{missing mass}: a yet-unseen type of matter distributed throughout the universe whose presence could explain several problematic astronomical observations (\citealt{Einasto1974}, \citealt{Ostriker1974}). The publication of these papers indicates the establishment of what is currently known as the `dark matter' problem -- one of the most well-known anomalies in the prevailing cosmological model. According to this model, 85\% of the universe's mass budget consists of dark matter. After four decades of multi-wavelength astronomical observations and high-energy particle physics experiments, the nature of this mass is yet to be determined.\footnote{For an overview of the physics, see e.g. \citealt{Bertone2005,Bertone2010,Bertone2018}.} 

The origin of the dark matter problem serves more than only to illustrate the persistence of this fascinating anomaly. In this paper, I argue that the early justification for the existence of dark matter lays bare the foundation and formation of the contemporary discipline of cosmology. 
In the original 1974 papers, the two research groups put together a series of earlier published observations and interpreted them as signaling the presence of unseen mass. Crucially, both papers reflected on the cosmological significance of their conclusions: the tenfold-increased mass of the universe they had found, agreed with a decades-old cosmological model in which the universe is `geometrically closed'. This model was ``believed strongly'' by physicists ``for essentially nonexperimental reasons'' (\citealt{Ostriker1974}: L1). Dark matter  mattered because it could accommodate such a model. 

The establishment of dark matter's existence is discussed in more detail in a previous paper (\citealt{DeSwart2017}). In the current paper, I explore the significance of the above argument. Where did this `closed universe' come from, how can we understand its legitimacy, and what does it tell us about the practice of cosmology and its history? I argue that the reasoning which helped to establish the dark matter problem is closely entangled with the maturation of the discipline of cosmology, from the 1950s to the 1970s. Specifically, understanding the way that dark matter came to matter reveals the development of post-war cosmology's methodological character. It illuminates the two faces of this character that emerged in the 1970s: a fruitful hybrid of principle-based deductive and observational-based empirical approaches -- two faces that still show in cosmological discussions today.\footnote{Contributions to methodological discussions on inflation, string theory and the multiverse often emphasize either empirical data or deductive thought. For this opposition see e.g. \citealt{Ellis2014}.} 

This paper builds on many authoritative works that have been written on the history of cosmology in the second half of the twentieth century.\footnote{For broader overviews of the history of cosmology in the second half of the twentieth century see in particular \cite{North1965}, \cite{Kragh1996}, \cite{Smeenk2003}, \cite{Kragh2006}, and \cite{Longair2013}.} These works have particularly focused on the theoretical developments that brought about the emergence of the big-bang cosmological paradigm in the early 1970s. The current paper, instead, centers on studying the changes in how cosmology was practiced. Where in the 1950s there was no unanimously celebrated theory, method or practice that defined what a science of the universe looked like, by the early 1970s there was an established theoretical and observational cosmological canon. I analyze this development by tracing the continuities and discontinuities in what it meant to do cosmology during this period.  

My take on the establishment and foundations of modern cosmology centers around understanding the rise of dark matter as an inescapable part of its history, and as an early exemplar of its methods.  I argue that dark matter's establishment particularly lays bare the hybrid character of modern cosmology: the dark matter problem arises in the application of a combination of methodological strategies, in which the roots of both  `rationalist' and `empiricist' approaches to cosmology reverberate. Dark matter's confluence of different methodological styles is traced back to two historical developments: the wake of general relativity theory's blooming development in the 1950s, and the wealth of observations and institutional changes that remade the astronomical sciences in the 1960s. 

The essay then is split up in four parts, and has four different goals: (1) illustrate the history of cosmology's methodological foundations in the 1950s; (2) discuss the relation between cosmology and the revival of research on general relativity in the late 1950s; (3) indicate the fundamental conceptual and institutional changes in the practice of astronomy during the 1960s, which signals cosmology's `rebirth'; and, (4) argue that by the 1970s, instead of a single cosmological method appearing victorious, cosmology synthesized different scientific norms and practices, as shown by the rise of the dark matter problem.


\section{The Foundations of Cosmology in the 1950s}
\label{sec:2}

Partially due to the fruits of post-war observational programs in astronomy, the early 1950s knew increased attention to cosmological issues. There were new estimates of the rate of expansion of the universe -- the Hubble constant, $H_0$ -- by Baade (1952) and Humason, Mayall and Sandage (1958); radio astronomers agreed on the extra-galactic nature of observed radio sources (\citealt{Ryle1956}); and, new catalogues of galaxies appeared from surveys of unprecedented size done by the Lick and Palomar observatories (\citealt{Shane1956,Abell1959}). Although extra-galactic scales started to be systematically explored (cf. \citealt{Smith2008a}), good and cosmologically relevant observations were still considered scarce and hard to obtain. Indeed, empirical novelties were only partially responsible for the increased attention to cosmological research in the 1950s. A large part of the increasing work and publications on cosmology in that period was due to a clash between two theories of the cosmos: relativistic cosmology and the steady state theory of the universe. Due to this clash, cosmology as a science was under severe scrutiny during the 1950s.

Relativistic cosmology was the approach rooted in Einstein's 1917 effort to treat the entire universe with his relativistic field equations. Although Einstein initially introduced a static cosmological model, relativistic cosmology was believed to imply an evolutionary model of the universe from the 1930s onward. This new dynamical interpretation owed its existence to the observational work of Milton Humason and Edwin Hubble, and theoretical developments initiated by Alexander Friedmann and Georges Lema\^itre. George Gamov had connected relativistic cosmology with nuclear physics in the 1940s, but, by the 1950s, relativistic cosmology was still mainly understood in terms of the Friedmann-Lemaitre picture: as a model (or set of models) in which the dynamics of Einstein's field equations describe the expansion of the universe. This model was the early stage of what today is understood as the big bang theory.\footnote{Note that the idea of an expanding universe did not intrinsically involve the hypothesis of a cosmic origin, or what was known as Lema\^itre's ``primeval-atom'' hypothesis. This idea was mainly celebrated by Lema\^itre and Gamov, but it was no integral part of relativistic cosmology at that time. Furthermore, note that the name ``relativistic cosmology'' is very much a convention. Some form of relativity was necessarily used in all cosmologies 
(cf. \citealt{McCrea1953}: 350). For a detailed expos\'e on early relativistic cosmology, see especially \cite{Kragh1996}.}

Criticism of the expanding relativistic model was hardly new, but it received a very concrete form in 1948 with the `steady state theory'.
The theory was introduced by Cambridge physicists Hermann Bondi and Thomas Gold (1948), and mathematician Fred Hoyle (1948), as an attempt to resolve a discrepancy between the relativistic theory and observations. This discrepancy mainly consisted of a mismatch between the time-scales of the relativistically predicted age of the universe, and the much larger estimated age of the Earth.\footnote{After the revision of the Hubble constant by Baade in 1952, the mentioned time-scale problem became less problematic for relativistic cosmology, although it did not fully disappear. See also: \cite{Bondi1952}, p. 140.} Bondi, Gold, and Hoyle also had serious reservations about the unphysical and unscientific nature of the idea of a cosmic origin, which, although not unanimously celebrated by relativists, was implied by an expanding universe. Their alternative was a universe that on the large scale is steady and unchanging. It was infinite of age with a constant average density. This solved both the time-scale difficulty and the need for a big bang. The steady state theorists notoriously hypothesized a ``continuous creation of matter'' to reconcile their ideas with a constant average density in an expanding universe. These ideas were met with skepticism by many relativists.  

Steady state theorists had challenged the physical foundations of relativistic cosmology. The resulting debate between these parties dominated cosmological practice in the 1950s. Just as Helge \cite{Kragh1996} has emphasized, these debates show the fundamental philosophical character of cosmological practice in the 1950s.\footnote{Helge Kragh has treated the details of these discussions in his 1996 book (\citealt{Kragh1996}: 219-251). Much of the tensions of the methodological debates in 1950s cosmology are rooted in early debates from the 1930s. See e.g.: \cite{Gale1999}. 
} At the same time, the explicit questioning of its foundation hints to cosmology's pre-paradigmatic stage as a discipline; there lacked a unanimously acclaimed theory, method or practice that defined what a science of the cosmos looked like. I explore these debates to particularly lay bare the methodological positions that form the roots of current-day cosmology.

\subsubsection*{``Empiricists'' and ``Rationalists''}

The very nature of cosmology as a proper science of the universe was at stake in the debate between steady state theory and relativistic cosmology. How do you study the cosmos, and what methods should guide this inquiry? Because of the severe friction between the groups of theorists, such philosophical questions started to be elaborately discussed in the 1950s. The different methodological positions in this debate were often divided into two camps by the involved theorists. 
Steady state theorist Bondi wrote about ``extrapolating'' versus ``deductive'' attitudes in cosmology (Bondi 1952: 5), whereas relativist George McVittie identified them as ``empiricist'' and ``rationalist'' approaches (\citealt{McVittie1961b}: 12). 
These two styles were argued to roughly coincide with the two theories of the cosmos: steady state theory had a deductive approach where relativistic cosmology was based on extrapolation of general relativity theory.\footnote{Many authors used different terminology for these two styles. William McCrea, for example, wrote about ``deductive'' and ``astrophysical'' attitudes in cosmology (\citealt{McCrea1953}: 332). 
}   We will see that a strict division in approaches might not do historical justice to all theorists, but the dichotomy does highlight the prevailing major methodological tensions.

The split between ``rationalist'' and ``empiricist'' approaches chiefly concerned what primacy one would award to physical principles. In 1932, Edward Milne formulated an extension of Einstein's famed principle of relativity -- the idea that all frames of reference know the same laws of nature. Milne's extended principle held that not only the laws of nature but the universe itself must appear to have the same structure to every observer (Milne 1932: 10). This idea has since been known as the `cosmological principle'. In more contemporary terms, the principle holds that the matter distribution in the universe is homogeneous and isotropic on large scales.\footnote{As with many physical concepts, the exact formulation of the cosmological principle differed between authors. Dennis Sciama formulated the cosmological principle as ``[e]ach particle always sees an isotropic distribution of particles around it'' (\citealt{Sciama1960}: 312). McVittie put it slightly differently:  ``[t]he development of the universe appears to be the same for each observer of an equivalent set, every one of whom assigns co-ordinates by the same method'' (\citealt{McVittie1952}: 96).} The cosmological principle quickly turned into a central element of attempts to formulate theories for the structure of the universe. However, the prominence given to the principle in building and justifying these theories differed widely.\footnote{In a review of cosmology in 1953, McCrea wrote: ``All the theories to be discussed require their models to conform to the cosmological principle (CP), though we shall see that they do so for somewhat different reasons'' (\citealt{McCrea1953}:326).}
That is, should the principle serve as an addendum to theoretical exploration, or does this principle form the very condition of possibility of a science of the universe?

For Bondi and Gold, the cosmological principle explicitly served the latter goal. In their 1948 paper, they reasoned that the cosmological principle is a logical necessity for the universe to be intelligible at all, as it protects the universal applicability of physical laws. Given that physical laws are tested locally, Bondi and Gold argued that there is no reason to assume that they could not differ throughout the universe.
Then, in pursuing a science of the cosmos, one should provide grounds on which it can actually be assumed that the local physical laws hold everywhere the same. A system of cosmology should be principally concerned with this, Bondi and Gold wrote; it should be able to justify the ``unrestricted repeatability of all experiments'' (\citealt{Bondi1948}: 252). For them, relativistic cosmology lacked a way to guarantee this unrestricted repeatability. 

The steady state theorists found the grounds on which to justify the universal application of known physical laws in what they called the ``perfect cosmological principle'' (\textit{idem}: 254). Their extension of the cosmological principle said that the large-scale universe looks the same on every position in both space \textit{and} time. To avoid any dependencies of physical laws, in future and past, one should not only assume that the universe looks the same everywhere, but also that it is unchanging. Hence, as Bondi put it, one should ``postulate that position in space and time is irrelevant'' (\citealt{Bondi1952}: 11). Without the perfect cosmological principle, ``cosmology'' they wrote, ``is no longer a science'' (\citealt{Bondi1948}: 255). Because of their primacy of \textit{a priori} principles, these steady steady theorists were identified as rationalist or deductivist.

For Hoyle's version of the steady state theory things worked a bit differently. He worked on a relativistic extension of Einstein's field equations, and did not share Bondi and Gold's worries about the validity of physical laws. Hoyle emphasized that formulations of these laws in terms of fields would mean that their validity is guaranteed everywhere. In a similar fashion he disagreed about the role of the perfect cosmological principle. He remarked in 1949: ``[i]t is believed that the wide [perfect] cosmological principle should follow as a consequence of primary axioms of the field form [...] and should not appear itself as a primary axiom'' (\citealt{Hoyle1949}: 371). 

Despite Hoyle's different attitude, it seems that we can  still make sense of the fact that he was often grouped with Bondi and Gold in having a deductive approach. This comes from the fact that Hoyle's justification of a steady state theory in 1948 hinged primarily upon the introduction of continuous creation of matter (\citealt{Hoyle1948}). This initial assumption was often criticized on being an ``arbitrary alteration'' of the field equations (Heckmann, in \citealt{Stoops1958}:76). There was no experimental evidence for such an alteration, and hence it was argued there was no reason to abandon other fundamental principles (e.g. energy conservation) because of this idea. Hoyle, involved in studies of stellar nucleosynthesis, was no less concerned with observations than were the proponents of relativistic cosmology. His disagreement with relativistic cosmology, however, does seem to have been considered a matter of principle.

The claim of having a deductivist or rationalist approach to cosmology was not restricted to steady state theorists. In the 1930s, Edward Milne's introduction of a kinematic relativistic world model -- a  description of the universe without a need for Einstein's theory -- was found to be based on a hypothetico-deductive method. Lema\^itre even traced rationalist philosophical attitudes in cosmology back to Leibniz.\footnote{See: \cite{Milne1935}. In 1958 Lema\^itre stated: ``As far as I can see, the inclination to rely on an a priori principle is related to Leibnitz philosophical attitude which made him to believe that there is some esthetical design in the Universe or even that the Universe is determined as being the best possible one'' (\citealt{Lemaitre1958}: 2). }  During the 1950s, however, this approach began to be discussed programmatically and more explicitly. This happened as steady state theorists used their methodological convictions to defend their theory. In his 1952 textbook on cosmology, Bondi wrote that ``it is a dangerous habit of the human mind to generalize and to extrapolate without noticing that it is doing so'' (\citealt{Bondi1952}: 6). Thomas Gold more specifically warned against uncritical extrapolation of known physical laws: ``[n]o prejudices about physical principles must be used there when they would be based merely on the acquaintance with a much smaller scale of the physical world'' (\citealt{Gold1956}: 1721). In these discussions, the dichotomy of approaches to cosmology started to take shape. 

The methodology behind relativistic cosmology was less explicitly discussed; it was formulated mainly as a response to the steady state theory. The cosmological principle was also less central in relativistic cosmology. The principle was used as an additional criterion for obtaining special solutions of the field equations, not as an axiomatic statement. Influential cosmologists in the 1950s advocating an approach to cosmology that opposed that of Bondi and Gold included McVittie, and astronomer and philosopher Herbert Dingle. Especially the latter was fierce in his response to the steady state theorists. In 1956, Dingle, former president of the Royal Astronomical Society, wrote the following: ``[s]ome cosmologists have returned to the discredited practice of inventing arbitrary general principles, with no justification except that they seem `right,' and fitting phenomena to the requirements of the principles'' (\citealt{Dingle1956}: 234). 

Most relativists were similarly, but much less aggressively, in pursuit of an empirical methodology. German cosmologist Engelbert Sch\"ucking and astronomer Otto Heckmann, for example, found that ``[a] theory constructed on a sound foundation of empirical data ought not to be discarded unless [...] new facts turn up that cannot be fitted into the framework of this theory'' (\citealt{Schucking1958}: 149). The theory to which Sch\"ucking and Heckmann refer here is Einstein's theory of general relativity. For relativistic cosmologists, the legitimacy of Einstein's theory was given priority over principles. When in the mid-1950s a revival of interest and trust in the potential of Einstein's theory commenced, this served as an offensive against the deductive approach.


\section{Relativity's Renaissance, Cosmology and Mach's Principle}

Although the flourishing modern era of gravitational wave astronomy and black hole physics would seem to suggest otherwise, research on the physics of gravity has been waxing and waning throughout the twentieth century. Whereas the subject bloomed shortly after Einstein's introduction of the theory of general relativity, historian Jean Eisenstaedt has shown that from the mid-1920s to the 1950s gravitational physics knew a low-water-mark period of stagnated research activity (\citealt{Eisenstaedt1986,Eisenstaedt1987,Eisenstaedt1989}). Only in the mid-1950s did the tide turn for gravity's relative weight in physics research. During a period characterized by community formation and the recognition of the general theory of relativity's untapped physical potential, an integrated research field of gravitational physics arose. 
Physicist Clifford Will dubbed this period the ``renaissance of general relativity'' (\citealt{Will1986,Will1989}).  

Research on gravity during the renaissance period knew an enormous shift to experimental and observational issues in relativity. These issues included, for example, the possible measurement of gravitational waves.\footnote{See \cite{Peebles2017} for an elaborate historical overview of the history of experimental gravitational physics. For more on the history of gravitational waves during that period, see \cite{Blum2018}.} With this experimental focus, and in combination with the usage of new mathematical tools, the renaissance period knew a vast exploration of general relativity as the fundamental theory of gravity. These developments during general relativity's renaissance have been elaborately discussed by \cite{Blum2015,Blum2016,Blum2017}, and I will follow their heuristic periodization.\footnote{See also e.g. \cite{Eisenstaedt2006}, and \cite{Lalli2017}.} 
What has been discussed in lesser extent, is how general relativity's developments reverberated in cosmological research.  The ``global transformation in the character of [general relativity]'' that Blum et al. have addressed (\citealt{Blum2017}: 98), further implied a transformation in the relationship between cosmology and relativity.  In particular, the increased authority of the general relativity theory during its renaissance emphasized a relativity-based extrapolating attitude to cosmological research. 

Steady state theorists, often trained in relativity, joined in meetings on gravitational physics, and brought the discussion on how to approach cosmology to the attention of gravity scholars. One of the most influential scholars in the new field of gravitational physics that responded to this issue was Princeton professor John Wheeler. During one of relativity's renaissance famous meetings -- the 1957 Chapel Hill conference -- he clearly expressed his view on how to approach cosmology. He commented on the work of steady state theorist Thomas Gold by stating that ``one should not give up accepted ideas of wide applicability such as general relativity but should investigate them completely'' (Wheeler, in \citealt{Rickles57}: 129). Wheeler's take on cosmology was the relativistic view that started to dominate the discussions during the renaissance of general relativity: an approach to cosmology that centered around the extrapolation of general relativity.

This view became even more ingrained as confidence in the potential of the theory of general relativity kept increasing. By the early 1960s, the general opinion was that general relativity was fundamentally true and that this needed little debate. Venerable physicist and Princeton colleague of Wheeler, Robert Dicke, wrote that ``[specialists] take it as axiomatic that general relativity is correct in all its details and that one must compute with this theory'' (\citealt{Dicke1964}: 1). That relativity was axiomatic also held for how cosmology was perceived to be done. In 1962, Wheeler reflected on the transformation that the status of general relativity had undergone:

\begin{quotation}
Increasing numbers of investigators share the conviction that Einstein's 1915-1916 analysis of the curvature of space by energy is a unique theory, of unrivalled scope and reasonableness, against which no objection of principle or observation has ever been sustained, and out of which one should now try to read the deeper meaning and consequences. Among these consequences some of the most interesting have to do with the dynamics of the universe. (\citealt{Wheeler1962}: 40)
\end{quotation}

Between about 1955 and 1963, the relation between cosmology and relativity transformed. Where Einstein's former research assistant Peter Bergmann wrote in 1957 that cosmology was ``not intimately connected'' with other aspects of relativity (\citealt{Bergmann1957}: 352), this sentiment had changed by the early 1960s.  General relativity was a supplying a new ``rationale'' for practicing cosmology, as Wheeler put it (\citealt{Wheeler1962}: 75).   The increasing confidence in the theory of relativity made cosmological research relevant and urgent: cosmology became the very consequence and testing grounds for the theory of general relativity. 

General relativity as a rationale for cosmology meant that, in the early 1960s, the climate changed around the steady state versus big bang theory debate. As the steady state theory indirectly challenged general relativity through its dismissal of relativistic cosmology, the latter began to be the more favored one. Helge Kragh similarly reflected that relativistic cosmology's increasing status is ``probably related to the simultaneous revival of interest in the general theory of relativity'' (\citealt{Kragh1996}: 318). He also pointed out that Hoyle's, McCrea's, and later versions of the steady state theory were being designed to conform to the mathematics of general relativity through Einstein's Field Equations, which again emphasizes the gravitational theory's increasing authority  (\citealt{Kragh1999}: 398). 

The `empiricist' approach became the dominant way to address the method of cosmology. But how `empirical' was this approach really? According to McVittie, avid defender of relativistic cosmology, the general relativity point of view was ``that scientific cosmology should be based on the laws of physics as we know them from experiment and observation rather than on hypotheses and principles laid down a priori'' (\citealt{McVittie1961a}: 1232). However, as I hope will be clear, one could be quite skeptical of McVittie's remarks on the method of relativistic cosmology. That is, the less empirical and more \textit{philosophical} foundations of Einstein's theory itself also kept lingering within cosmology.

\subsubsection*{The Mach Connection}
A sharp reader might have already been aware of the perhaps confusing identification of relativistic cosmology as `empirical' in approach. The confusion comes from the fact that general relativity theory's methodological underpinning is much more elusive.\footnote{For a comprehensive overview of the genesis of general relativity, see \cite{Renn2007a}. For an in-depth discussion of Einstein's methodology and its development, see \cite{Dongen2010}.} 
``[T]he origins of General Relativity are mainly philosophical rather than observational'', Robert Dicke reflected in 1961 (\citealt{Dicke1962}: 4). One of these philosophical foundations has been specifically recognized: the idea known as Mach's principle. This principle was a persistent subject in discussions on general relativity when it accompanied the rebirth of interest in the theory during the 1950s.\footnote{Many elaborate studies have been done on Mach's Principle, its general importance and its role in the theory of general relativity. See specifically \cite{Barbour1995}. The writing of a longue dur\'ee history of the principle seems to have not yet been attempted.} Understanding the role of this principle helps to understand the ambiguous methodological character of cosmological practice, both in relativistic and steady state cosmology.

 Ernst Mach's famed principle is reminiscent of his elaborate critique on Newton's absolute notions of space and motion. Opposing Newton, Mach had argued that mechanics is not based on absolute, but on the relative position and movement of bodies. 
This relationist view had far-reaching implications for Mach's notion of inertia. Because there is no absolute rotation, he argued that an observer on the surface of a sphere should notice no physical difference in either having the sphere rotate with respect to distant objects, or having distant objects rotate around the sphere. However, we know that on the surface of a rotating sphere (e.g. the earth), one experiences inertial forces like that of the centrifugal force. With Mach's insights, ``the principles of mechanics'' could be so understood ``that even for relative rotations centrifugal forces arise'' (\citealt{Mach1960}: 284).  
 Because of this reasoning, Mach suggested there to be a relationship between local inertial forces, and the distant celestial bodies. This is what has come to be known as Mach's principle.\footnote{Einstein was the first who had formulated Mach's ideas on inertia as a ``principle'' (\citealt{Einstein1918}: 16).} 

Einstein was profoundly influenced by Mach's ideas on inertia; for a long time he was concerned with having a theory of gravity that fully abides by Mach's mandate.\footnote{\label{note1}For the role of Mach's principle in the development of general relativity, see e.g. \cite{Hoefer1994}, \cite{Renn2007b}, \cite{Barbour2007}, \cite{Lehmkuhl2014}, and \cite{janssen2014}.} As relativity revived in the 1950s, so did these concerns. The principle came to have a wider interpretation, and more generally was understood as the idea that the local inertial frame depends on, or, in a stronger form, is determined by the mass distribution of the universe. In this sense, the adjective `Machian' was used for any connection between local dynamics and the structure of the universe.\footnote{See \cite{Goenner1970} and references therein for examples of the use of `Mach's principle' and `Machian' in the 1950s and 1960s. There were many different and more technical definitions.} 
During the renaissance of general relativity, many august physicists took Mach's principle to be fundamental to gravity theories. Much research was done on how to make Einstein's field equations come in full accord with the principle and whether one could think of a true Machian theory of gravity. 

Dennis Sciama, who is often recognized as one of the fathers of modern cosmology, was central in the revival of work on Mach's principle. His doctoral work under the supervision of Paul Dirac resulted in a novel take on the  ``origin of inertia'', in which he proposed a type of long-range interaction with distant matter (\citealt{Sciama1953}). Mach's principle was put high on the list of the newly developing research agenda of gravitational physics. Referring to Sciama's work, Wheeler orated at the Chapel Hill conference in 1957 that one of general relativity's important open problems was ``spelling out Mach's principle in a better-defined way'' (\citealt{Wheeler1957}: 46). Robert Dicke was similarly concerned with Mach's ideas. He wrote about its importance on many occasions in the late 1950s and early 1960s, and was influential in discussing experiments on testing Mach's principle through possible mass anisotropies in the universe.\footnote{See e.g. \cite{RobertDicke1959}, \cite{Dicke1962} and for experimental test also \cite{Cocconi1960}, and \cite{Dicke1961}. For a historical overview of Dicke's important work, see \cite{Peebles2017}.} In 1961, he and Carl Brans proposed what came to be known the Brans-Dicke theory of gravitation, to come into accordance with Machian ideas (\citealt{Brans1961}).

In cosmology too, Mach's principle was of great importance during the 1950s. It served as a foundation for both steady state and relativistic cosmology. Mach's principle formed a central motivation for the steady state theorists' `Perfect Cosmological Principle'. That is, when there exists a Machian connection between the cosmic distribution of mass and local inertia, a different mass distribution in the earlier universe might imply different local laws of physics. The steady state theorists tried to avoid such a possibility with their reasoning.\footnote{E.g. Bondi wrote that ``[f]or in any theory which contemplates a changing universe, explicit and implicit assumptions must be made about the interactions between distant matter and local physical laws. These assumptions are necessarily of a highly arbitrary nature, and progress on such a basis can only be indefinite and uncertain. [...] If the uniformity of the universe is sufficiently great none of these difficulties arise'' (\citealt{Bondi1952}: 12).} 

Relativistic cosmology's relation with Mach's principle had deeper and more complex historical ties. In Einstein's cosmological work of 1917, he again emphasized that inertia should follow from a body's gravitational interaction with all the other masses in the universe. In any model of the universe, he hence reasoned, a mass at sufficient distance from all the other masses should have its inertia fall to zero. Einstein tried different ways of implementing this Machian condition. He found it most satisfactory to use it as a selection criterion: the relativistic model that can realize the Machian condition is the correct one. For Einstein, this model was a universe that is spatially closed.\footnote{As both Chris \cite{Smeenk2014} and Carl \cite{Hoefer1995} have clearly spelled out, Einstein first tried to use Mach's statement as a boundary condition to the field equations. Later, in his 1917 cosmology paper, he had turned away from this perspective. Instead, he used the fact that a spatially closed universe has no boundary region. Einstein noted: ``[f]or if it were possible to regard the universe as a continuum which is finite (closed) with respect to its spatial dimensions, we should have no need at all of any such boundary conditions'' (\citealt{einstein1987}: 427). For an in-depth discussion of Einstein's 1917 paper, see \cite{ORaifeartaigh2017}.} Einstein later more specifically reflected on his epistemological preference for a closed universe: 

\begin{quotation}
[T]his idea of Mach's corresponds only to a finite universe, bounded in space [...]. From the standpoint of epistemology it is more satisfying to have the mechanical properties of space completely determined by matter, and this is the case in a closed universe. (\citealt{Einstein1922}: 108)
\end{quotation}

Whether or not adding the assumption of a closed universe would make general relativity truly Machian was, and still is, disputed.\footnote{For more information on whether general relativity is Machian see e.g. \cite{Dicke1962}, and more generally \cite{Barbour1995} and references found in footnote \ref{note1}.} What cannot be disputed is that the preference for a closed universe became wide-spread. %
Having a universe that is closed was even seen as an integral part of Einstein's theory of relativity. In 1958, Wheeler considered the following to be the meaning of the term ``Einstein's Relativity'':

\begin{quotation}
In speaking about Einstein's theory [...] we mean not only the system of differential equations associated with his name, but also two further points, the present tentative arguments for which he gives in his book [\textit{The Meaning of Relativity} (1922)]: 

(1) The universe is closed; 

(2) No ``cosmological'' term is to be added to the field equations. (\citealt{Wheeler1958}: 98)
\end{quotation}

Although by the early 1960s general relativity appeared ``victorious'' in cosmology, as Kragh has put it (\citealt{Kragh1999}: 398), this did not mean the empiricist approach had swept away every \textit{a priori} principle or rationalistic tendency. Given that general relativity was a `theory of principle' by its very origin, there was, on a fundamental level, no approach that can be regarded as a clear-cut winner in cosmological methodology. On the surface, general relativity was applied cosmologically in an empiricist manner, but underneath, relativity had many `rationalistic' features that echoed through in cosmology. These features were decreasingly explicit from the mid-1960s onward. From that moment, the efforts of relativity, astrophysics and cosmology became dominated by newly observed phenomena.


%
\section{The Golden Age: a Cosmological Turn}
\label{sec:3}
The gained trust in the theory of general relativity manifested itself in what Kip \cite{Thorne1994} has called a `golden age' for relativity: from 1963 onward, the theory became largely celebrated because of its applications to astronomy and astrophysics. This golden age got much of its stature due to a wealth of newly observed astronomical phenomena, and immense conceptual changes in the understanding of the universe's structure and contents. During this period, research on cosmology again became re-characterized. It became less driven by relativists and was increasingly integrated into the interests of astronomers and astrophysicists, which stressed an observational research program in cosmology. I will give three examples of these changes of interests. 

Firstly, as Robert \cite{Smith2008a} has also argued, astronomy increasingly shifted toward extra-galactic phenomena. New catalogues of galaxies based on the surveys of the Palomar and Lick observatories in the late 1950s made it possible to better understand the large scale distribution of galaxies.\footnote{These catalogues include \cite{Zwicky1961}, \cite{Vorontsov-VelYaminov1962}, and \cite{Arp1966}.}  This included, for example, the conclusions that \textit{clusters} of galaxies were a fundamental part of this distribution.\footnote{In 1957, George Abell wrote that ``[p]rior to 1949, only a few dozen clusters were known. [...] In recent years, however, two independent photographic programs have indicated that clusters of galaxies are far more numerous than was formerly thought, and that indeed they may be fundamental condensations of matter in the universe'' (\citealt{Abell1957a}: 3).} 
The advent of radio astronomy in the 1950s had similarly influenced extra-galactic astronomy; it opened up a wavelength region that spanned more than a factor of 1000. The Cambridge C3 radio catalogues supplied astronomers with a wealth of new data that led to the observation of unidentified phenomena in galaxies and the universe: radio luminosities that are orders of magnitude larger than their optical counterparts, and objects like double radio sources, radio stars, and radio galaxies.\footnote{For the radio catalogues see \cite{Edge1959}, and \cite{Bennett1962}. See also \cite{Sciama1971a}, pp. 49-82.} 
 
The great post-war sky surveys were accompanied with new instrumentation that similarly influenced the turn to extra-galactic astronomy. In 1961, Hubble's successor at Mt. Wilson observatory, Allan Sandage, had published an optimistic piece on how the 200-inch Hale telescope, that first saw light in 1949, could potentially discriminate between different cosmological models. He wrote: ``[r]enewed interest in the cosmological problem is evidenced by the number of recent papers which treat the fitting of observational data to predictions of the theory'' (\citealt{Sandage1961}: 356). Sandage's analysis of possible cosmological tests and available observations revitalized a program in observational cosmology.

The second illustration of the golden age's conceptual changes is how relativity became a genuine subject of concern for astronomers and astrophysicists. This was connected to an observation that was particularly influential in making the 1960s a golden age for general relativity: the first observation of a quasar by Maarten Schmidt in 1963. The brightness of this radio object was 100 times larger than any known galaxy, implying an explosive energy release. The unknown nature of this quasar acted as a boundary problem that brought astronomy and gravitational physics close together. 

As an immediate response to the quasar discovery, the first of a very successful series of ``Texas Symposia on Relativistic Astrophysics'' was held in Dallas, Texas. Here astronomers, astrophysicists, cosmologists, gravitational physicists, and nuclear physicists participated to discuss the nature of these quasars. The main motivation for the symposium was indeed the energy related to the quasars: it was calculated to be more than $10^{60}$ ergs, which is the energy contained in the rest mass of a million solar masses (cf. \cite{Sciama1971a}: 61). The theoretical requirements needed for an outburst of such enormous energy had ``so far ruled out nearly all of the explanations and theories put forward to explain such extraordinary events'', the invitation to the symposium read (\citealt{Robinson1965}: v). 

Gravitational physicists were involved in searching for a mechanism of gravitational collapse that might cause such outbursts of energy. Or, as John Wheeler noted at the Texas symposium, ``attention has turned to gravitational collapse as a mechanism by which in principle a fraction of the latent energy of matter much larger than 1 percent can be set free''  (\citealt{Harrison1965}: 1). Proposed mechanisms included the formation of objects that we now understand as black holes and neutron stars. The conference illustrates that the quasar observation had indeed brought physicists and astronomers very close together. At the dinner of the First Texas Symposium, Thomas Gold, of steady state fame, made this quite clear: ``[e]veryone is pleased, the relativists [...] who are suddenly experts in a field they hardly knew existed; the astrophysicists for having enlarged their domain, their empire, by the annexation of another subject -- general relativity'' (\citealt{Gold1965}: 470).

The third point that demonstrates changing concerns in the 1960s, was the increasing focus on questions of origin and evolution of the universe. Quasars were the oldest objects astronomers had observed, and because of their age, they started to be used to track the evolution of galaxies. Their age gave insight in the early evolutionary state of galaxies, and how galactic properties change with time. In the 1960s, the topic of galaxy evolution began to be explored. ``The study of evolution of galaxies is now in an early stage of development comparable to that of the study of stellar evolution in 1935'', Thornton Page reflected in 1964 (\citealt{Page1964a}: 804). Better knowledge of galactic evolution meant that galaxies could be used as probes for the evolution of kinematic and gravitational properties of the universe. The evolutionary picture of galaxies emphasized what the study of galaxies could mean for cosmology; the initial conditions of galaxies are closely tied to the contents and evolution of the universe.

The above examples are far from a complete overview of the flood of phenomena that entered astronomical research in the 1960s. Also included in this list are the observation of the cosmic microwave background in 1964, and the first pulsar observation in 1967. Both these observations had an enormous theoretical impact, but a detailed discussion of this impact would go beyond the purpose of this paper. The given examples suffice to show how the focus of astronomers and astrophysicist was rapidly changing in the 1960s. The flood of observations had swamped the earlier philosophical reflections, in favor of a  ``renaissance in observational cosmology'', as Dennis Sciama called it (\citealp{Sciama1971b}). 
Indeed, in 1970, the National Research Council reported to U.S. Congress that ``the rapid pace of discovery in astronomy and astrophysics during the last few years has given this field an excitement unsurpassed in any other area of the physical sciences'' (\citealt{NationalResearchCouncil1972}: 55).

The dominance of observations in the 1960s redetermined the objects of interest for both physicists and astronomers, and blurred the boundaries between physics and astronomy. This is also illustrated by the manpower distribution in astronomy.


\subsubsection*{Manpower and Textbooks}
\label{sec:4}

The renewed interest of astronomers and physicists was accompanied by large transformations in the institutional landscape of astronomy in the 1960s. Specifically, physicists were starting to flood astronomy in the late 1960s and early 1970s. By the early 1970s, astronomy had many more, and very different practitioners compared to earlier decades. 

In the U.S., the 1960s knew an enormous increase in astronomy graduate students, quite probably related to the explosive demands of the space program. The number of degrees awarded in astronomy had increased tenfold by 1970, compared to a decade earlier. 
The number of awarded astronomy degrees grew at an accelerated rate: where between 1920 and 1960 the number of astronomy Ph.D.-degrees awarded grew around 4\% every year, between 1960 and 1970 this rose to a 20\% annual increase. Not only the number of students increased. The total number of personnel employed in the field of astronomy, both technical and scientific, almost tripled in the 1960s. \footnote{\cite{NationalResearchCouncil1973}, p. 327.} 

Besides an absolute increase in manpower, the background of people that worked in astronomy also changed. The field became increasingly dominated by physicists. While in 1966, 26 percent of the astronomy personnel with Ph.D.s had received their doctorates in physics, by 1970 this had increased to 45 percent. In 1970 it was projected that it took only two more years until there were more people with Ph.D.s in physics working in astronomy, than people with Ph.D.s in astronomy.
\footnote{\cite{NationalResearchCouncil1972}, pp. 55-56; \cite{NationalResearchCouncil1973}, p. 332.} Although the funding channels were stagnating in astronomy, the discipline was doing relatively well compared to physics in the early 1970s; in 1971 the unemployment rate of Ph.D.s in physics was more than four times as high as in astronomy.\footnote{\cite{NationalResearchCouncil1973}, p. 337. David \cite{Kaiser2002} has in great detail written about the case of manpower in American physics after World War II.}  This could explain part of the large influx of physicists into astronomy.

Indeed, the institutional character of astronomy had changed by the early 1970s, and physics seemed to have taken a larger place within the discipline. At the same time, the subject of cosmology was becoming increasingly popular. 
The use of ``cosmologist'' as profession began to circulate in the 1960s (\citealt{Kragh2006}: 200), and the number of publications on cosmology increased with an order of magnitude between 1965 and 1975 (\citealt{DeSwart2017}: 4). 
Furthermore, with the increasing number of students, many new textbooks on cosmology appeared in the early 1970s:  Jim Peebles' ``Physical Cosmology'' (1971); Dennis Sciama's ``Modern Cosmology'' (1971); Weinberg's ``Gravitation and Cosmology'' (1972); Hawking and Ellis' ``The Large Scale Structure of Space-Time'' (1973); and, Thorne, Misner and Wheeler's ``Gravitation'' (1973).\footnote{More examples of textbooks are Robertson and Noonan's ``Relativity and Cosmology'' (1968),  Wolfgang Rindler's ``Essential Relativity: Special, General, and Cosmological'' (1969), and Zeldovich and Novikov's ``Relativistic Astrophysics: Stars and Relativity'' (1971).}  These books again show the boundary-crossing character of cosmology: every single author of these books was a physicist by training, and many of them discuss astronomical observations, cosmological theory, and gravitational physics.

Where the mid-1960s was dominated by the renaissance of observational cosmology, the early 1970s, perhaps with the influence of physicists, theoretical cosmology also flourished. Backed by radio star counts and the cosmic microwave background, relativistic cosmology had become the canon cosmological theory, in which different models could be explored. There was a practical way in which theoretical cosmology also started to flourish. Between 1967 and 1970, 32\% of the astronomer Ph.D. recipients found that their research was restricted by the lack of availability of observing time on local and national telescopes.\footnote{\cite{NationalResearchCouncil1973}, p. 349.} While observational capacities were lacking, theoretical possibilities were widely available. By 1970, `theoretical astrophysics' was the research subject that enjoyed most the interest of professional astronomers, according to the National Research Council.\footnote{\cite{NationalResearchCouncil1972}, p. 60.} 

The boundary between physics and astronomy was blurring and astronomical research underwent a cosmological turn: research was primed towards the study of the universe. Cosmological research formed a hybrid environment that blended extra-galactic astronomy, nuclear physics, astrophysics, and gravitation physics. This new \textit{physical} cosmology, as it was sometimes referred to, was accommodated by astronomy's increased institutional scale and inflow of physicists. This again had consequences for how cosmology was perceived to be done.


\section{Birth and Rebirth: Cosmology and the Dark Matter Problem}
\label{sec:5}

By the 1970s, the status of cosmological research had radically transformed. Cosmology now was deeply driven by observations and had become a commonplace subject that crossed the boundaries between physics and astronomy. Following \cite{Merleau-Ponty1976}, it seems we can justly signify this transformation as a ``Rebirth of Cosmology''.\footnote{\cite{Merleau-Ponty1976} seem to use two interpretations of ``The Rebirth of Cosmology'', the title of their book: either as one of both cosmological revolutions instigated by Newton and Einstein, or in the sense of the narrower period in which cosmology had transformed to be the frontier of science by 1976. I use it in the latter sense.} Quite similar to  what happened in the 1950s to general relativity -- as a theory, and as a field of research -- cosmology had become a respectable part of the physical sciences.

Cosmology's rebirth is most visible in how its status as a `proper' subject of study became undisputed. 
The achievements of cosmology, ``especially in the last few years'', Dennis Sciama wrote in 1971, ``constitute a revolution in our knowledge and understanding of the Universe with no parallel in the whole recorded history of mankind'' (\citealt{Sciama1971a}: vii). Where, in the 1950s, renowned radio astronomer Martin Ryle had been notoriously skeptical of the whole cosmological enterprise,\footnote{In the 1950s Ryle noted ``[c]osmologists always lived in a happy state of being able to postulate theories which had no chance of being disproved [...]'' (Ryle quoted in \citealt{Kragh1996}: 309).} he was awarded the Nobel Prize in 1974 for his radio-astronomical work and invention of the aperture synthesis technique, that had been of ``crucial significance'' for cosmology.\footnote{``Press Release: The 1974 Nobel Prize in Physics''. Nobelprize.org. Nobel Media AB 2014. Web. 30 Jan 2018. For more on the curious relationship between the Nobel prize and the astronomical sciences, see \cite{kragh2017}. 
}  It was the first time the prestigious prize was awarded for astronomical research. Ryle's student, Malcolm Longair, portrayed the shift in the status of cosmology:

\begin{quotation}
Because of the increased confidence in the hot [Big Bang] model and the larger number of real facts about the universe, topics which in the past were questions of pure speculation have become susceptible to detailed quantitative analyses which can be checked against the observations. (\citealt{Longair1971}:1125-1248)
\end{quotation}

The speculative days were over, and little traces were left of the great philosophical discussions of the 1950s. Early textbooks like that of \cite{McVittie1956}  had vividly discussed the nature of scientific laws, and \cite{Bondi1952} devoted whole sections to the cosmological principle, the problem of inertia, and the differences between physics and cosmology. Textbooks of the early 1970s avoided such elaborate philosophy. Instead, the observational potential of cosmology was celebrated, and textbooks and monographs exhibited a shift to an implicit, but familiar `empiricist' style of reasoning. As Steven Hawking and George Ellis, both students of Sciama, wrote in their 1973 textbook: ``we shall take the local laws of physics that have been experimentally determined, and shall see what these laws imply about the large scale structure of the universe'' (\citealt{Hawking1973}: 1).

Another example of the overwhelming dominance of the `empiricist' approach is Dennis Sciama's celebrated oeuvre on cosmology. While in the 1950s, Sciama wrote a popular book that was themed around Mach's principle -- titled ``The Unity of the Universe'' (\citealp{Sciama1959}) -- his later book 
barely mentions Mach. Instead, he emphasized that ``[t]he first question must be: can [the great flood of new observations] be understood in terms of the known laws of physics?'' (\citealt{Sciama1971a}: 101). Mach's principle was considered in great extent by pre-1960s textbooks, but publications from the 1970s had mostly left these discussions aside.

The new generation of relativists and cosmologists were barely exposed to the philosophical underpinnings of cosmology. The tendency among the new pupils in gravitation and cosmology was clear: the phenomena came first, not the principles.  Yet the vivid methodological discussions of the 1950s did have their repercussions in cosmological research of the 1970s. Similar to how cosmology became a hybrid environment of astronomers and physicists, a fruitful hybrid of approaches to cosmology came into use. Stylistic remnants of deductive approaches can be recognized in the a priori preferences that existed for certain cosmological models.  

\subsubsection*{A Closed Universe}

In the reborn science of the cosmos, there were many different models in which the theory of the explosive universe could be realized. From the Friedmann equations follow three different cases for the evolution of the universe. These were given by the value of the curvature of the universe: negative (spatially open), positive (spatially closed), or a special model which had no intrinsic curvature (spatially 'flat').  This spatial curvature was indirectly measurable, and in the early 1970s there was optimism about being able to distinguish between these models observationally. 

``[B]eginning in the 1960s a flood of new discoveries has enriched our picture of the universe and has begun to provide a basis on which to distinguish between competing cosmological models'', Allen Sandage wrote in 1970 (\citealt{Sandage1970}: 34). For him, cosmology was a ``search for two numbers'': the Hubble constant, and the deceleration parameter. The latter, in practice, was directly related to the mass density of the universe.\footnote{In a universe without a cosmological constant, $q_0$ and $\rho$ are directly related by the equation $\rho/\rho_c = 2 q_0$, with $\rho_c$ the critical density.} This parameter became a central research subject in cosmology and extra-galactic astronomy. The density of the universe relates to its destiny, and measuring it could determine which of the cosmological models corresponds to the universe: open or closed. The density was often expressed in terms of $\Omega$, the density relative to the `critical density': $\Omega = \rho / \rho_{critical}$. The critical density ($\Omega = 1$) was the density of a universe that was `flat', which was a model introduced by \cite{Einstein1932}. Less mass would mean an open universe ($\Omega < 1$), more mass would mean a spatially closed universe ($\Omega > 1$). 

Although Sandage, as an observer, indeed emphasized that observations will determine which model is correct, preconceptions about the shape of the universe still lingered. As discussed in Section 2, Mach's principle had left a strong imprint on the way relativistic cosmology was done. The principle was ``conceived as the requirement that the universe be closed'', as Wheeler had put it just before the `Golden Age' of relativity took off (\citealt{Wheeler1962}: 74). Nearing the end of the 1960s, a closed universe was still a much-preferred model for many cosmologists and relativists. ``\textit{Philosophically}, there might be a preference'', Wolfgang Rindler wrote in 1967, ``the choice $k=1$ [a positively curved universe] might appear desirable. It implies closed space sections that would, in some sense, validate Mach's principle according to which the totality of matter in the universe and nothing else determines the local inertial frames'' (\citealt{Rindler1967}: 29-30, emphasis in original). Dennis Sciama had a different preference, but similar reasoning for preferring a universe with a specific density: 

\begin{quotation}
[...] the Einstein-de Sitter model is the one where the total energy of the universe is zero, the kinetic energy and the negative gravitational potential energy just balancing. Well, if you think that kinetic energy manifesting inertia is due to gravitation, then you might intuit that the most Machian way of having one made by the other would be if there's equal amount of energy, which would give you uniquely the Einstein-de Sitter model, I still have a secret hope that that might turn out so, but it may well not. (Sciama, in \citealp{Weart1978})
\end{quotation}

The implicit preference for a closed universe was also expressed in observational studies. ``One would particularly like to know whether there is enough mass to close the universe'', Princeton physicists Peebles and Partridge wrote in 1967, in a piece on estimating the mass density of the universe (\citealt{Peebles1967}: 713). However, there was a crucial discrepancy between such observations and a spatially closed cosmological model: the measured value of the mass density of the universe was typically around the order of $10^{-31}~gr.~cm^{-3}$, two orders of magnitude lower than the mass needed to close the universe.\footnote{See e.g. \cite{Oort1958}; \cite{Peebles1971}; \cite{Shapiro1971}; \cite{Noonan1971}; \cite{Weinberg1972}, p. 478; \cite{Burbidge1972}, p. 493. The critical density is $\rho_c = \frac{3 H^2}{8\pi G} \sim 10^{-29}~gr.~cm^{-3}$.} In his 1972 textbook, Steven Weinberg wrote: 

\begin{quotation}
[I]f one tentatively accepts the result that $q_0$ is of order unity [$\Omega \geq 1$], then one is forced to the conclusion that the mass density of about $2 \times 10^{-29}g/cm^3$ must be found somewhere outside the normal galaxies. But where? (\citealt{Weinberg1972}: 478)
\end{quotation}

The idea of a closed universe still found resonance with the older generation of relativists and cosmologists, like John \cite{Wheeler1974}. But also the newer generation of cosmologists worried about the implication of this ``theological'' idea that the universe ought to be closed:\footnote{J. Richard Gott received a Ph.D. in 1973, Jim Gunn in 1966, David Schramm in 1971, and Beatrice Tinsley in 1966.} 

\begin{quotation}
Where [can] the missing mass be hiding if it is demanded, on theological or other grounds that $\Omega \geq 1 \space$ [$\rho\geq\rho_c$]. (\citealt{GottJ.R.1974}: 550)
\end{quotation}

The discrepancy between the observed mass density and the density needed for a closed universe made that a new problem appeared: the ``problem of the so-called `missing mass''', as Geoffrey Burbidge called it; the mass missing to close the universe (\citealt{Burbidge1972}: 493). Such mass would ultimately vindicate Mach's principle that was so vigorously pursued in earlier decades, but this reason behind preferring a closed universe did seldom enter in discussions. Although its origin was little regarded, the philosophical preference for extra mass did have real implications for astrophysical and cosmological observational programs.  Research was set to uncover potential yet-unseen intergalactic matter and to explore how to  theoretically accommodate such a finding.\footnote{In a 1972 review, George Field wrote that ``[t]he main interest in IGM [inter-galactic matter] stems from the evidence that galactic matter constitutes only a small fraction of the critical cosmological density of matter and energy [...]'' (\citealt{Field1972}: 227-228).} An answer to the demand for extra mass came in 1974, from two independent collaborations of astronomers and physicists.

\subsubsection*{The Birth of Dark Matter}
Although cosmology had received the status of an empirical science in the early 1970s, there  still lingered part of a `rationalist' character in its practice. To have a closed universe -- the cosmological model initially inspired by Ernst Mach -- would mean a density much higher than was determined by observations of the luminosity of stars and galaxies. The dark matter problem finds its birth in this context, where two observations were recognized as indicating the existence of this `missing mass'. 

The first of these observations concern the dynamics of galaxies. Much of the wealth of data on galaxies that was acquired in the 1960s and early 1970s was related to radio astronomy. Radio astronomical studies of galaxies were done in great number, with new telescopes like the Owens Valley Radio Observatory interferometer and the Westerbork Radio Synthesis Telescope. With radio astronomy, galactic properties could be measured much beyond a galaxy's luminous radius. One of these measured properties was the rotational velocity of galaxies. By the early 1970s it was found that these rotation curves had a peculiar characteristic: the velocity of rotation stayed ``flat'', i.e. constant out to large distances where there was no mass to account for this velocity.\footnote{For examples of these radio astronomical studies, see e.g. \cite{Roberts1973,Rogstad1972,Rogstad1973}. Influential optical studies were also done by \cite{Freeman1970} and \cite{Rubin1970}.} 

Around the same time, clusters of galaxies were found to be fundamental to the universe's structure, and a related anomaly that was first reported in the 1930s regained attention. In clusters of galaxies, the masses of individual galaxies did not add up to make sense of the observed dynamical state of the cluster. That is, the cluster’s observed stability could not be explained only by the visible mass.\footnote{Important overviews of the problematic dynamics of clusters include \cite{Neyman1961}, \cite{Page1967}, \cite{Rood1970} and \cite{Field1971}. The cluster problem was first remarked by Fritz \cite{Zwicky1933}.} Both the flat rotation curves and the cluster discrepancy were well-known curiosities in the late 1960s and early 1970s. Only in 1974, however, these observations were linked together and interpreted as signaling the existence of missing mass. This was done by two independent research groups.  

In their 1974 article, the Princeton collaboration of astronomer Jerry Ostriker, and physicists Jim Peebles and Amos Yahil incorporated the observations mentioned above into a single argument. They connected different distance scales on which masses of galactic systems were dynamically calculated, to make an estimate of the mass density of the universe. Their analysis showed that the observations corresponded to a linear increase of mass with radius. The interest of the authors to bring these observations of galactic masses together was profoundly cosmological, and their conclusion even more so. Their introduction read as follows: 

\begin{quotation}
There are reasons, increasing in number and quality, to believe that the masses of ordinary galaxies may have been underestimated by a factor of 10 or more. [...] the current estimate (Shapiro 1971) for the ratio of gravitational energy to kinetic energy in the universe is about $\Omega = 0.01$. If we increase the estimated mass of each galaxy by a factor well in excess of 10, we increase the ratio by the same amount and conclude that observations may be consistent with a Universe which is  ``just closed'' ($\Omega = 1$) -- a conclusion believed strongly by some (cf. Wheeler 1973) for essentially nonexperimental reasons. (\citealt{Ostriker1974}: L1)
\end{quotation}

In the establishment of the dark matter problem, the demand for extra mass acted as a motivation to put together already existing observations and issues of mass, and create a picture consistent with a closed universe. The paper of Wheeler to which the authors referred, again emphasized the attractive power of a closed universe, and  ``the mystery of the missing mass'' that it entailed (\citealt{Wheeler1974}: 686).\footnote{Note that it was even the case that a tenfold increase in galactic masses for the authors meant that ``observations may be consistent'' with a hundredfold increase in the universe mass density (from 0.01 to 1).} As discussed in Section 2, this belief in a closed universe was a remnant of Einstein's epistemological considerations concerning general relativity back in the beginning of the twentieth century.

The argument given by Ostriker et al. shows how cosmological practice was still embedded in the philosophical roots of cosmology and relativity.  In the early 1970s, cosmology was born anew, fed by observations and the testability of its models. But the formation of this cosmological paradigm all but meant that a single approach to cosmology was practiced. Dark matter, in this sense, exemplifies the hybridity of the reborn cosmology: hybrid in the sense of its participants, being both astronomers and physicists; hybrid in its matters of concern, ranging from scales of galaxies to clusters; and, most illustratively, hybrid in its methodology, combining detailed observations with a priori conceptions of the cosmos. The rationale for this belief was no longer explicitly discussed as observational programs had become dominant, but, in practice, a closed universe kept being an influential and fruitful part of cosmological research.  

Two months before the Princeton group, a similar argument was published by astronomer Jaan Einasto, cosmologist Ants Kaasik and Enn Saar from Tartu Observatory in Estonia in the USSR. They similarly frame their results in terms of the critical density needed to close the universe:

\begin{quotation}
Evidence is presented that galaxies are surrounded by massive coronas exceeding the masses of known stars by one order of magnitude [...] the total density of matter in the galaxies being 20\% of the critical cosmological density. (\citealt{Einasto1974a}: 2)
\end{quotation}

Only with these two publications in 1974, the series of observations that are now considered evidence for dark matter were unambiguously interpreted as signaling the existence of yet-unseen mass; observations that acquired their meaning in the context of an age-old a priori consideration. 


\section{Conclusion}

Between 1950 and 1970 the field of cosmology revolutionized. The cosmological program that was initiated by Einstein in 1917 truly materialized during this post-war period. Where in the early 1950s cosmology was dominated by meta-scientific discussions, the field had established itself as a respectable empirical science by 1970. This rebirth of cosmology was kick-started by a renovation in physics: the renaissance of the theory of general relativity between 1955 and 1963. The promise of general relativity theory was rediscovered during that era and this reverberated in its application to cosmology. General relativity's renaissance was followed by a period in which an overwhelming number of cosmologically relevant observations were done, and cosmology's ambiguous methodological foundations seemingly fell into oblivion. By the early 1970, there was a newly established theoretical and observational cosmological canon; cosmology was born anew as a genuine physical science. 

Although they are often discussed independently from one another, I argue that the rise of the dark matter problem in 1974 is most illustrative of cosmology's new paradigmatic shape. It lays bare four aspects of cosmology's rebirth. (I) It shows how cosmology had become a hybrid field of both physicists and astronomers: the two landmark papers that put the dark matter problem on the map were both due to such an interdisciplinary collaboration. (II) It demonstrates how the flood of new astronomical observations appearing in the 1960s fed into cosmological research: dark matter's claim to fame was built on two observations that are reminiscent of this decade of observational abundance. (III) Perhaps more tentatively, the rise of dark matter illustrates how modern cosmological practice involves drawing together a vast amount of different scales that are relevant for understanding the evolution and structure of the universe: the dark matter problem rose from data on galactic rotation curves and the extra-galactic dynamics of galaxy clusters. (IV) The 1974 papers highlight that multiple methodological strategies underlie cosmological practice. In dark matter's case, anomalous observations were made relevant by relating them to the potential spatial closure of the universe; a model that was preferred purely on \textit{a priori} basis.  

The latter point has been central to my paper. Although in the 1970s cosmology became a respected subject to be studied empirically, it remained richly permeated with extra-theoretical considerations. A hybrid of `rationalist' and `empiricist' methods came to characterize cosmological practice after its mid-century rebirth. I argue that to understand the practice of modern cosmology, we must forgo normative claims on which approach is the correct one, and instead accept that cosmology's methodology is fundamentally plural in character. Cosmology is a ``methodological omnivore''; rather than focusing on a single method or test, the field drives on the convergence of multiple evidential means. This notion is used by philosopher Adrian Currie to describe the historical sciences (\citealt{Currie2015}), and the analysis presented in the current paper seems a case-in-point to understand cosmology in a similar fashion. 

Cosmology's omnivorous character again surfaces in research on contemporary hot topics. Particularly in discussions on the merits of inflation theory, the almost dialectic tension between empirical and deductive strategies in cosmology resurfaces, as is clear in a recent controversy published in Scientific American (\citealt{Ijjas2017}). Similarly, non-empirical arguments are still widely used in discussions surrounding the multiverse (cf. \citealt{Kragh2009}). But also within empirical programs different observational strategies are combined: the search for dark matter recently was argued to enter a new era ``with the new guiding principle being `no stone left unturned''' (\citealt{Bertone2018}: 54). Perhaps this then is a lesson to draw from cosmology's history, and the way the dark matter problem was brought to light: scientific knowledge of the universe is not acquired by a single approach, but by a hybrid of different methods.

\begin{acknowledgement}
I am most grateful to Helge Kragh, Sjang ten Hagen and the editors of this volume, J\"urgen Renn, Alexander Blum and Roberto Lalli, for helpful comments on an earlier draft of this paper. Many thanks go also to Jeroen van Dongen and Gianfranco Bertone for their guidance, suggestions and inspiring discussions on the project.
\end{acknowledgement}

\bibliographystyle{humlinphil-contrib} 
\bibliography{Bibliography}

\begin{thebibliography}{}

\bibitem[\protect\astroncite{Abell}{1957}]{Abell1957a}
Abell, G.~O. (1957).
\newblock {\em {The Distribution of Rich Clusters of Galaxies}}.
\newblock California Institute of Technology.

\bibitem[\protect\astroncite{Abell}{1959}]{Abell1959}
Abell, G.~O. (1959).
\newblock {The National Geographic Society-Palomar Observatory Sky Survey}.
\newblock {\em Astronomical Society of the Pacific Leaflets}, 8(366):121.

\bibitem[\protect\astroncite{Arp}{1966}]{Arp1966}
Arp, H. (1966).
\newblock {Atlas of Peculiar Galaxies}.
\newblock {\em The Astrophysical Journal Supplement Series}, 14:1.

\bibitem[\protect\astroncite{Barbour}{2007}]{Barbour2007}
Barbour, J.~B. (2007).
\newblock {\em {Einstein and Mach's Principle}}, pages 1492--1527.
\newblock Springer Netherlands, Dordrecht.

\bibitem[\protect\astroncite{Barbour and Pfister}{1995}]{Barbour1995}
Barbour, J.~B. and Pfister, H., editors (1995).
\newblock {\em {Mach's Principle: from Newton's Bucket to Quantum Gravity}}.
\newblock Birkhauser, Boston.

\bibitem[\protect\astroncite{Bennett and Smith}{1962}]{Bennett1962}
Bennett, A.~S. and Smith, F.~G. (1962).
\newblock {The Preparation of the Revised 3C Catalogue of Radio Sources}.
\newblock {\em Monthly Notices of the Royal Astronomical Society},
  125(1):75--86.

\bibitem[\protect\astroncite{Bergmann}{1957}]{Bergmann1957}
Bergmann, P.~G. (1957).
\newblock {Summary of the Chapel Hill Conference}.
\newblock {\em Reviews of Modern Physics}, 29(3):352--354.

\bibitem[\protect\astroncite{Bertone}{2010}]{Bertone2010}
Bertone, G., editor (2010).
\newblock {\em {Particle Dark Matter: Observations, Models and Searches}}.
\newblock Cambridge University Press, Cambridge.

\bibitem[\protect\astroncite{Bertone et~al.}{2005}]{Bertone2005}
Bertone, G., Hooper, D., and Silk, J. (2005).
\newblock {Particle Dark Matter: Evidence, Candidates and Constraints}.
\newblock {\em Physics Reports}, 405(5-6):279--390.

\bibitem[\protect\astroncite{Bertone and Tait}{2018}]{Bertone2018}
Bertone, G. and Tait, T. M.~P. (2018).
\newblock {A new era in the search for dark matter}.
\newblock {\em Nature}, 562(7725):51--56.

\bibitem[\protect\astroncite{Blum et~al.}{2017}]{Blum2017}
Blum, A., Giulini, D., Lalli, R., and Renn, J. (2017).
\newblock {Editorial introduction to the special issue “The Renaissance of
  Einstein's Theory of Gravitation”}.
\newblock {\em The European Physical Journal H}, 42(2):95--105.

\bibitem[\protect\astroncite{Blum et~al.}{2015}]{Blum2015}
Blum, A., Lalli, R., and Renn, J. (2015).
\newblock {The Reinvention of General Relativity: A Historiographical Framework
  for Assessing One Hundred Years of Curved Space-time}.
\newblock {\em Isis}, 106(3):598--620.

\bibitem[\protect\astroncite{Blum et~al.}{2018}]{Blum2018}
Blum, A., Lalli, R., and Renn, J. (2018).
\newblock {Gravitational waves and the long relativity revolution}.
\newblock {\em Nature Astronomy}, 2(7):534--543.

\bibitem[\protect\astroncite{Blum et~al.}{2016}]{Blum2016}
Blum, A.~S., Lalli, R., and Renn, J. (2016).
\newblock {The renaissance of General Relativity: How and why it happened}.
\newblock {\em Annalen der Physik}, 528(5):344--349.

\bibitem[\protect\astroncite{Bondi}{1952}]{Bondi1952}
Bondi, H. (1952).
\newblock {\em {Cosmology}}.
\newblock Cambridge University Press, Cambridge.

\bibitem[\protect\astroncite{Bondi and Gold}{1948}]{Bondi1948}
Bondi, H. and Gold, T. (1948).
\newblock {The Steady-State Theory of the Expanding Universe}.
\newblock {\em Monthly Notices of the Royal Astronomical Society},
  108(3):252--270.

\bibitem[\protect\astroncite{Brans and Dicke}{1961}]{Brans1961}
Brans, C. and Dicke, R.~H. (1961).
\newblock {Mach's Principle and a Relativistic Theory of Gravitation}.
\newblock {\em Physical Review}, 124(3):925--935.

\bibitem[\protect\astroncite{Burbidge}{1972}]{Burbidge1972}
Burbidge, G.~R. (1972).
\newblock {Intergalactic Matter and Radiation (Survey Lecture)}.
\newblock In Evans, D.~S., Wills, D., and Wills, B.~J., editors, {\em External
  Galaxies and Quasi-Stellar Objects, Proceedings from IAU Symposium no. 44
  held in Uppsala, Sweden, 10-14 August 1970}. D. Reidel, Dordrecht.

\bibitem[\protect\astroncite{Cocconi and Salpeter}{1960}]{Cocconi1960}
Cocconi, G. and Salpeter, E.~E. (1960).
\newblock {Upper Limit for the Anisotropy of Inertia from the M{\"{o}}ssbauer
  Effect}.
\newblock {\em Physical Review Letters}, 4(4):176--177.

\bibitem[\protect\astroncite{Currie}{2015}]{Currie2015}
Currie, A. (2015).
\newblock {Marsupial lions and methodological omnivory: function, success and
  reconstruction in paleobiology}.
\newblock {\em Biology {\&} Philosophy}, 30(2):187--209.

\bibitem[\protect\astroncite{de~Swart et~al.}{2017}]{DeSwart2017}
de~Swart, J.~G., Bertone, G., and van Dongen, J. (2017).
\newblock {How dark matter came to matter}.
\newblock {\em Nature Astronomy}, 1(3):0059.

\bibitem[\protect\astroncite{Dicke}{1959}]{RobertDicke1959}
Dicke, R.~H. (1959).
\newblock {New Research on Old Gravitation}.
\newblock {\em Science}, 129:621--624.

\bibitem[\protect\astroncite{Dicke}{1961}]{Dicke1961}
Dicke, R.~H. (1961).
\newblock {Experimental Tests of Mach's Principle}.
\newblock {\em Physical Review Letters}, 7(9):359--360.

\bibitem[\protect\astroncite{Dicke}{1962}]{Dicke1962}
Dicke, R.~H. (1962).
\newblock {Mach's Principle and Equivalence}.
\newblock In M{\o}ller, C., editor, {\em Proceedings of the International
  School of Physics Enrico Fermi XX: Evidence for Gravitational Theories. Held
  between June 19 and July 1 in Verenna, Italy.}, page 264. Academic Press.

\bibitem[\protect\astroncite{Dicke}{1964}]{Dicke1964}
Dicke, R.~H. (1964).
\newblock {Remarks on the Observational Basis of General Relativity}.
\newblock In Chiu, H.-Y. and Hoffman, W.~F., editors, {\em Gravitation and
  relativity}, pages 1--16. W.A. Benjamin, Inc, New York.

\bibitem[\protect\astroncite{Dingle}{1956}]{Dingle1956}
Dingle, H. (1956).
\newblock {Cosmology and Science}.
\newblock {\em Scientific American}, 195(3):224--240.

\bibitem[\protect\astroncite{Edge et~al.}{1959}]{Edge1959}
Edge, D.~O., Shakeshaft, J.~R., McAdam, W.~B., Baldwin, J.~E., and Archer, S.
  (1959).
\newblock {A survey of radio sources at a frequency of 159 Mc/s.}
\newblock {\em Memoirs of the Royal Astronomical Society}, 68:37--60.

\bibitem[\protect\astroncite{Einasto et~al.}{1974a}]{Einasto1974}
Einasto, J., Kaasik, A., and Saar, E. (1974a).
\newblock {Dynamic evidence on massive coronas of galaxies}.
\newblock {\em Nature}, 250(5464):309--310.

\bibitem[\protect\astroncite{Einasto et~al.}{1974b}]{Einasto1974a}
Einasto, J., Kaasik, A., and Saar, E. (1974b).
\newblock {Dynamical evidence on massive coronas of galaxies}.
\newblock {\em Tartu Astrof{\"{u}}{\"{u}}s. Obs. Preprint}, 1:8.

\bibitem[\protect\astroncite{Einstein}{1918}]{Einstein1918}
Einstein, A. (1918).
\newblock {Prinzipielles zur allgemeinen Relativit{\"{a}}tstheorie}.
\newblock {\em Annalen der Physik}, 360(4):241--244.

\bibitem[\protect\astroncite{Einstein}{1922}]{Einstein1922}
Einstein, A. (1922).
\newblock {\em {The Meaning of Relativity}}.
\newblock Springer Netherlands, Dordrecht.

\bibitem[\protect\astroncite{Einstein}{1987}]{einstein1987}
Einstein, A. (1987).
\newblock {Cosmological Considerations in the General Theory of Relativity}.
\newblock In {\em The Collected Papers of Albert Einstein. Volume 6: The Berlin
  Years: Writings, 1914-1917 (English translation supplement)}, pages 421--431.
  Princeton University Press, Princeton, NJ.

\bibitem[\protect\astroncite{Einstein and de~Sitter}{1932}]{Einstein1932}
Einstein, A. and de~Sitter, W. (1932).
\newblock {On the Relation between the Expansion and the Mean Density of the
  Universe}.
\newblock {\em Proceedings of the National Academy of Sciences}, 18:1--2.

\bibitem[\protect\astroncite{Eisenstaedt}{1986}]{Eisenstaedt1986}
Eisenstaedt, J. (1986).
\newblock {La relativit{\'{e}} g{\'{e}}n{\'{e}}rate {\'{a}} l'{\'{e}}tiage:
  1925 - 1955}.
\newblock {\em Archive for History of Exact Sciences}, 35:115--185.

\bibitem[\protect\astroncite{Eisenstaedt}{1987}]{Eisenstaedt1987}
Eisenstaedt, J. (1987).
\newblock {Trajectoires et Impasses de la Solution de Schwarzschild}.
\newblock {\em Archive for History of Exact Sciences}, 37:275--357.

\bibitem[\protect\astroncite{Eisenstaedt}{1989}]{Eisenstaedt1989}
Eisenstaedt, J. (1989).
\newblock {The Low Water Mark of General Relativity, 1925 - 1955}.
\newblock In {D. Howard} and {John Stachel}, editors, {\em Einstein and the
  History of General Relativity}, pages 277--292. Birkh{\"{a}}user, Boston.

\bibitem[\protect\astroncite{Eisenstaedt}{2006}]{Eisenstaedt2006}
Eisenstaedt, J. (2006).
\newblock {\em {The Curious History of Relativity: How Einstein's Theory of
  Gravity Was Lost and Found Again}}.
\newblock Princeton University Press, Princeton, NJ.

\bibitem[\protect\astroncite{Ellis and Silk}{2014}]{Ellis2014}
Ellis, G. and Silk, J. (2014).
\newblock {Scientific method: Defend the integrity of physics.}
\newblock {\em Nature}, 516(7531):321--3.

\bibitem[\protect\astroncite{Field}{1972}]{Field1972}
Field, G.~B. (1972).
\newblock {Intergalactic Matter}.
\newblock {\em Annual Review of Astronomy and Astrophysics}, 10(1):227--260.

\bibitem[\protect\astroncite{Field and Saslaw}{1971}]{Field1971}
Field, G.~B. and Saslaw, W.~C. (1971).
\newblock {Groups of Galaxies: Hidden Mass or Quick Disintegration?}
\newblock {\em The Astrophysical Journal}, 170:199.

\bibitem[\protect\astroncite{Freeman}{1970}]{Freeman1970}
Freeman, K.~C. (1970).
\newblock {On the Disks of Spiral and S0 Galaxies}.
\newblock {\em Astrophysical Journal}, 160:811.

\bibitem[\protect\astroncite{Gale and Urani}{1999}]{Gale1999}
Gale, G. and Urani, J. (1999).
\newblock {Milne, Bondi, and the 'Second Way' to Cosmology}.
\newblock In Goenner, H., Renn, J., Ritter, J., and Sauer, T., editors, {\em
  The Expanding Worlds of General Relativity}, pages 343--375.
  Birkh{\"{a}}user, Boston.

\bibitem[\protect\astroncite{Goenner}{1970}]{Goenner1970}
Goenner, H. (1970).
\newblock {\em {Mach's Principle and Einstein's Theory of Gravitation}}, pages
  200--215.
\newblock Springer Netherlands, Dordrecht.

\bibitem[\protect\astroncite{Gold}{1956}]{Gold1956}
Gold, T. (1956).
\newblock {Cosmology}.
\newblock {\em Vistas in Astronomy}, 2:1721--1726.

\bibitem[\protect\astroncite{Gold}{1965}]{Gold1965}
Gold, T. (1965).
\newblock {After-Dinner Speech}.
\newblock In Robinson, I., Schild, A., and Schucking, E.~L., editors, {\em
  Quasi-Stellar Sources and Gravitational Collapse: Proceedings of the 1st
  Texas Symposium on Relativistic Astrophysics}, page 470. University Of
  Chicago Press, Chicago.

\bibitem[\protect\astroncite{{Gott, J. R.} et~al.}{1974}]{GottJ.R.1974}
{Gott, J. R.}, I., Gunn, J.~E., Schramm, D.~N., and Tinsley, B.~M. (1974).
\newblock {An Unbound Universe}.
\newblock {\em The Astrophysical Journal}, 194:543.

\bibitem[\protect\astroncite{Harrison et~al.}{1965}]{Harrison1965}
Harrison, B.~K., Thorne, K.~S., Wakano, M., and Wheeler, J.~A. (1965).
\newblock {\em {Gravitation Theory and Gravitational Collapse}}.
\newblock niversity of Chicago Press, Chicago.

\bibitem[\protect\astroncite{Hawking and Ellis}{1973}]{Hawking1973}
Hawking, S.~W. and Ellis, G. F.~R. (1973).
\newblock {\em {The large-scale structure of space-time.}}
\newblock Cambridge University Press, Cambridge.

\bibitem[\protect\astroncite{Hoefer}{1994}]{Hoefer1994}
Hoefer, C. (1994).
\newblock {Einstein's struggle for a Machian gravitation theory}.
\newblock {\em Studies in History and Philosophy of Science Part A},
  25(3):287--335.

\bibitem[\protect\astroncite{Hoefer}{1995}]{Hoefer1995}
Hoefer, C. (1995).
\newblock {Einstein's Formulations of Mach's Principle}.
\newblock In Barbour, J.~B. and Pfister, H., editors, {\em Mach's Principle:
  from Newton's Bucket to Quantum Gravity}, pages 67--90. Birkhauser, Boston.

\bibitem[\protect\astroncite{Hoyle}{1948}]{Hoyle1948}
Hoyle, F. (1948).
\newblock {A New Model for the Expanding Universe}.
\newblock {\em Monthly Notices of the Royal Astronomical Society},
  108(5):372--382.

\bibitem[\protect\astroncite{Hoyle}{1949}]{Hoyle1949}
Hoyle, F. (1949).
\newblock {On the Cosmological Problem}.
\newblock {\em Monthly Notices of the Royal Astronomical Society},
  109(3):365--371.

\bibitem[\protect\astroncite{Ijjas et~al.}{2017}]{Ijjas2017}
Ijjas, A., Steinhardt, P.~J., and Loeb, A. (2017).
\newblock {Pop Goes the Universe}.
\newblock {\em Scientific American}, 316(2):32--39.

\bibitem[\protect\astroncite{Janssen}{2014}]{janssen2014}
Janssen, M. (2014).
\newblock {“No Success Like Failure...”}.
\newblock In Janssen, M. and Lehner, C., editors, {\em The Cambridge Companion
  to Einstein}, pages 167--227. Cambridge University Press, Cambridge.

\bibitem[\protect\astroncite{Kaiser}{2002}]{Kaiser2002}
Kaiser, D. (2002).
\newblock {Cold War requisitions, scientific manpower, and the production of
  American physicists after World War II}.
\newblock {\em Historical Studies in the Physical and Biological Sciences},
  33(1):131--159.

\bibitem[\protect\astroncite{Kragh}{1996}]{Kragh1996}
Kragh, H. (1996).
\newblock {\em {Cosmology and Controversy: The Historical Development of Two
  Theories of the Universe}}.
\newblock Princeton University Press.

\bibitem[\protect\astroncite{Kragh}{1999}]{Kragh1999}
Kragh, H. (1999).
\newblock {Steady State Cosmology and General Relativity: Reconciliation or
  Conflict?}
\newblock In Goenner, H., Renn, J., Ritter, J., and Sauer, T., editors, {\em
  The Expanding Worlds of General Relativity}, pages 377--402.
  Birkh{\"{a}}user, Boston.

\bibitem[\protect\astroncite{Kragh}{2006}]{Kragh2006}
Kragh, H. (2006).
\newblock {\em {Conceptions of cosmos: from myths to the accelerating universe:
  a history of cosmology}}.
\newblock Oxford University Press, Oxford.

\bibitem[\protect\astroncite{Kragh}{2009}]{Kragh2009}
Kragh, H. (2009).
\newblock {Contemporary History of Cosmology and the Controversy over the
  Multiverse}.
\newblock {\em Annals of Science}, 66(4):529--551.

\bibitem[\protect\astroncite{Kragh}{2017}]{kragh2017}
Kragh, H. (2017).
\newblock {The Nobel Prize System and the Astronomical Sciences}.
\newblock {\em Journal for the History of Astronomy}, 48(3):257--280.

\bibitem[\protect\astroncite{Lalli}{2017}]{Lalli2017}
Lalli, R. (2017).
\newblock {\em {Building the General Relativity and Gravitation Community
  During the Cold War}}.
\newblock SpringerBriefs in History of Science and Technology. Springer
  International Publishing.

\bibitem[\protect\astroncite{Lehmkuhl}{2014}]{Lehmkuhl2014}
Lehmkuhl, D. (2014).
\newblock {Why Einstein did not believe that general relativity geometrizes
  gravity}.
\newblock {\em Studies in History and Philosophy of Science Part B: Studies in
  History and Philosophy of Modern Physics}, 46:316--326.

\bibitem[\protect\astroncite{Lemaitre}{1958}]{Lemaitre1958}
Lemaitre, G. (1958).
\newblock {The Primeval Atom Hypothesis and the problem of the Clusters of
  Galaxies}.
\newblock In Stoops, R., editor, {\em La structure et l'{\'{e}}volution de
  l'univers}, pages 1--31. Institut International de Physique Solvay,
  Bruxelles.

\bibitem[\protect\astroncite{Longair}{1971}]{Longair1971}
Longair, M.~S. (1971).
\newblock {Observational cosmology}.
\newblock {\em Reports on Progress in Physics}, 34(3):306.

\bibitem[\protect\astroncite{Longair}{2013}]{Longair2013}
Longair, M.~S. (2013).
\newblock {\em {The Cosmic Century}}.
\newblock Cambridge University Press, Cambridge.

\bibitem[\protect\astroncite{Mach}{1960}]{Mach1960}
Mach, E. (1960).
\newblock {\em {The Science of Mechanics: A Critical and Historical Account of
  Its Development}}.
\newblock The Open Court Publishing Company, La Salle, Illinois, 6 edition.

\bibitem[\protect\astroncite{McCrea}{1953}]{McCrea1953}
McCrea, W.~H. (1953).
\newblock {Cosmology}.
\newblock {\em Reports on Progress in Physics}, 16(1):308.

\bibitem[\protect\astroncite{McVittie}{1952}]{McVittie1952}
McVittie, G.~C. (1952).
\newblock {\em {Cosmological Theory}}.
\newblock Methuen {\&} Co. Ltd., London, 2nd edition.

\bibitem[\protect\astroncite{McVittie}{1956}]{McVittie1956}
McVittie, G.~C. (1956).
\newblock {\em {General Relativity and Cosmology}}.
\newblock Chapman {\&} Hall, London.

\bibitem[\protect\astroncite{McVittie}{1961a}]{McVittie1961b}
McVittie, G.~C. (1961a).
\newblock {\em {Fact and Theory in Cosmology}}.
\newblock Macmillan, New York.

\bibitem[\protect\astroncite{McVittie}{1961b}]{McVittie1961a}
McVittie, G.~C. (1961b).
\newblock {Rationalism versus Empiricism in Cosmology}.
\newblock {\em Science}, 133(3460):1231--1236.

\bibitem[\protect\astroncite{Merleau-Ponty and
  Morando}{1976}]{Merleau-Ponty1976}
Merleau-Ponty, J. and Morando, B. (1976).
\newblock {\em {The Rebirth of Cosmology}}.
\newblock Alfred A. Knopf, New York.

\bibitem[\protect\astroncite{Milne}{1935}]{Milne1935}
Milne, E.~A. (1935).
\newblock {\em {Relativity, Gravitation and World-Structure}}.
\newblock The Clarendon Press, Oxford.

\bibitem[\protect\astroncite{{National Research
  Council}}{1972}]{NationalResearchCouncil1972}
{National Research Council} (1972).
\newblock {\em {Astronomy and Astrophysics for the 1970's: Volume 1: Report of
  the Astronomy Survey Committee}}.
\newblock National Academies Press, Washington, DC.

\bibitem[\protect\astroncite{{National Research
  Council}}{1973}]{NationalResearchCouncil1973}
{National Research Council} (1973).
\newblock {\em {Astronomy and Astrophysics for the 1970's: Volume 2: Report of
  the Panels}}.
\newblock National Academies Press, Washington, D.C.

\bibitem[\protect\astroncite{Neyman et~al.}{1961}]{Neyman1961}
Neyman, J., Page, T., and Scott, E. (1961).
\newblock {Conference on the Instability of Systems of Galaxies (Santa Barbara,
  California, August 10-12, 1961): Summary of the conference}.
\newblock {\em The Astronomical Journal}, 66(10):633.

\bibitem[\protect\astroncite{Noonan}{1971}]{Noonan1971}
Noonan, T.~W. (1971).
\newblock {The Mean Cosmic Density from Galaxy Counts and Mass Data}.
\newblock {\em Publications of the Astronomical Society of the Pacific}, 83:31.

\bibitem[\protect\astroncite{North}{1965}]{North1965}
North, J.~D. (1965).
\newblock {\em {The Measure of the Universe: A History of Modern Cosmology}}.
\newblock Clarendon Press, Oxford.

\bibitem[\protect\astroncite{Oort}{1958}]{Oort1958}
Oort, J.~H. (1958).
\newblock {Distribution of Galaxies and the Density in the Universe}.
\newblock In Stoops, R., editor, {\em La structure et l'{\'{e}}volution de
  l'univers}, pages 163--183. Institut International de Physique Solvay,
  Bruxelles.

\bibitem[\protect\astroncite{O'Raifeartaigh et~al.}{2017}]{ORaifeartaigh2017}
O'Raifeartaigh, C., O'Keeffe, M., Nahm, W., and Mitton, S. (2017).
\newblock {Einstein's 1917 static model of the universe: a centennial review}.
\newblock {\em The European Physical Journal H}, 42(3):431--474.

\bibitem[\protect\astroncite{Ostriker et~al.}{1974}]{Ostriker1974}
Ostriker, J.~P., Peebles, P. J.~E., and Yahil, A. (1974).
\newblock {The size and mass of galaxies, and the mass of the universe}.
\newblock {\em Astrophysical Journal}, 193:L1.

\bibitem[\protect\astroncite{Page}{1964}]{Page1964a}
Page, T. (1964).
\newblock {The Evolution of Galaxies}.
\newblock {\em Science}, 146(3645):804--809.

\bibitem[\protect\astroncite{Page}{1967}]{Page1967}
Page, T. (1967).
\newblock {Masses of galaxies: singles and members of multiple systems}.
\newblock In {\em Proceedings of the Fifth Berkeley Symposium on Mathematical
  Statistics and Probability}, volume~3.

\bibitem[\protect\astroncite{Peebles}{1971}]{Peebles1971}
Peebles, P. J.~E. (1971).
\newblock {\em {Physical Cosmology}}.
\newblock Princeton University Press, Princeton, NJ.

\bibitem[\protect\astroncite{Peebles}{2017}]{Peebles2017}
Peebles, P. J.~E. (2017).
\newblock {Robert Dicke and the naissance of experimental gravity physics,
  1957–1967}.
\newblock {\em The European Physical Journal H}, 42(2):177--259.

\bibitem[\protect\astroncite{Peebles and Partridge}{1967}]{Peebles1967}
Peebles, P. J.~E. and Partridge, R.~B. (1967).
\newblock {Upper Limit on the Mean Mass Density due to Galaxies}.
\newblock {\em The Astrophysical Journal}, 148:713.

\bibitem[\protect\astroncite{Renn}{2007a}]{Renn2007a}
Renn, J., editor (2007a).
\newblock {\em {The Genesis of General Relativity}}.
\newblock Boston Studies in the Philosophy of Science. Springer Netherlands,
  Dordrecht.

\bibitem[\protect\astroncite{Renn}{2007b}]{Renn2007b}
Renn, J. (2007b).
\newblock {\em {The Third Way to General Relativity: Einstein and Mach in
  Context}}, pages 945--1000.
\newblock Springer Netherlands, Dordrecht.

\bibitem[\protect\astroncite{Rickles and DeWitt-Morette}{1957}]{Rickles57}
Rickles, D. and DeWitt-Morette, C. (1957).
\newblock {The Role of Gravitation in Physics}.
\newblock In {\em Report from the 1957 Chapel Hill Conference}. Edition Open
  Sources.

\bibitem[\protect\astroncite{Rindler}{1967}]{Rindler1967}
Rindler, W. (1967).
\newblock {Relativistic Cosmology}.
\newblock {\em Physics Today}, 20(11):23--31.

\bibitem[\protect\astroncite{Roberts and Rots}{1973}]{Roberts1973}
Roberts, M.~S. and Rots, a.~H. (1973).
\newblock {Comparison of Rotation Curves of Different Galaxy Types}.
\newblock {\em Astronomy and Astrophysics}, 26:483--485.

\bibitem[\protect\astroncite{Robinson et~al.}{1965}]{Robinson1965}
Robinson, I., Schild, A., and Schucking, E.~L., editors (1965).
\newblock {\em {Quasi-Stellar Sources and Gravitational Collapse: Proceedings
  of the 1st Texas Symposium on Relativistic Astrophysics}}.
\newblock The University of Chicago Press, Chicago.

\bibitem[\protect\astroncite{Rogstad and Shostak}{1972}]{Rogstad1972}
Rogstad, D.~H. and Shostak, G.~S. (1972).
\newblock {Gross Properties of Five Scd Galaxies as Determined from 21-cm
  Observations}.
\newblock {\em The Astrophysical Journal}, 176:315.

\bibitem[\protect\astroncite{Rogstad et~al.}{1973}]{Rogstad1973}
Rogstad, D.~H., Shostak, G.~S., and Rots, A.~H. (1973).
\newblock {Aperture synthesis study of neutral hydrogen in the galaxies NGC
  6946 and IC 342.}
\newblock {\em Astronomy and Astrophysics}, 22:111--119.

\bibitem[\protect\astroncite{Rood et~al.}{1970}]{Rood1970}
Rood, H.~J., Rothman, V. C.~A., and Turnrose, B.~E. (1970).
\newblock {Empirical Properties of the Mass Discrepancy in Groups and Clusters
  of Galaxies}.
\newblock {\em The Astrophysical Journal}, 162:411.

\bibitem[\protect\astroncite{Rubin and Ford}{1970}]{Rubin1970}
Rubin, V.~C. and Ford, W.~Kent, J. (1970).
\newblock {Rotation of the Andromeda Nebula from a Spectroscopic Survey of
  Emission Regions}.
\newblock {\em The Astrophysical Journal}, 159:379.

\bibitem[\protect\astroncite{Ryle}{1956}]{Ryle1956}
Ryle, M. (1956).
\newblock {Radio Galaxies}.
\newblock {\em Scientific American}, 195(3):204--221.

\bibitem[\protect\astroncite{Sandage}{1961}]{Sandage1961}
Sandage, A.~R. (1961).
\newblock {The Ability of the 200-inch Telescope to Discriminate Between
  Selected World Models.}
\newblock {\em The Astrophysical Journal}, 133:355.

\bibitem[\protect\astroncite{Sandage}{1970}]{Sandage1970}
Sandage, A.~R. (1970).
\newblock {Cosmology: A search for two numbers}.
\newblock {\em Physics Today}, 23(2):34--41.

\bibitem[\protect\astroncite{Schucking and Heckmann}{1958}]{Schucking1958}
Schucking, E.~L. and Heckmann, O. (1958).
\newblock {World Models}.
\newblock In Stoops, R., editor, {\em La structure et l'{\'{e}}volution de
  l'univers}, pages 149--159. Institut International de Physique Solvay,
  Bruxelles.

\bibitem[\protect\astroncite{Sciama}{1953}]{Sciama1953}
Sciama, D.~W. (1953).
\newblock {On the Origin of Inertia}.
\newblock {\em Monthly Notices of the Royal Astronomical Society},
  113(1):34--42.

\bibitem[\protect\astroncite{Sciama}{1959}]{Sciama1959}
Sciama, D.~W. (1959).
\newblock {\em {The Unity of the Universe}}.
\newblock Faber and Faber, London.

\bibitem[\protect\astroncite{Sciama}{1960}]{Sciama1960}
Sciama, D.~W. (1960).
\newblock {Observational aspects of cosmology}.
\newblock {\em Vistas in Astronomy}, 3:311--328.

\bibitem[\protect\astroncite{Sciama}{1971a}]{Sciama1971a}
Sciama, D.~W. (1971a).
\newblock {\em {Modern Cosmology}}.
\newblock Cambridge: University Press, Cambridge, 1 edition.

\bibitem[\protect\astroncite{Sciama}{1971b}]{Sciama1971b}
Sciama, D.~W. (1971b).
\newblock {The recent renaissance of observational cosmology}.
\newblock In Kuper, C.~G. and Peres, A., editors, {\em Relativity and
  gravitation, Based on the proceedings of an International Seminar on
  Relativity and Gravitation, held at Technion City, Israel, July, 1969.}, page
  283. Gordon {\&} Breach, New York.

\bibitem[\protect\astroncite{Shane and Wirtanen}{1954}]{Shane1956}
Shane, C.~D. and Wirtanen, C.~A. (1954).
\newblock {The distribution of extragalactic nebulae}.
\newblock {\em Astronomical Journal}, 59:285.

\bibitem[\protect\astroncite{Shapiro}{1971}]{Shapiro1971}
Shapiro, S.~L. (1971).
\newblock {The Density of Matter in the Form of Galaxies}.
\newblock {\em The Astronomical Journal}, 76:291.

\bibitem[\protect\astroncite{Smeenk}{2003}]{Smeenk2003}
Smeenk, C. (2003).
\newblock {\em {Approaching the Absolute Zero of Time: Theory Development in
  Early Universe Cosmology}}.
\newblock PhD thesis, University of Pittsburgh.

\bibitem[\protect\astroncite{Smeenk}{2014}]{Smeenk2014}
Smeenk, C. (2014).
\newblock {Einstein's Role in the Creation of Relativistic Cosmology}.
\newblock In Janssen, M. and Lehner, C., editors, {\em The Cambridge Companion
  to Einstein}, pages 228--269. Cambridge University Press, Cambridge.

\bibitem[\protect\astroncite{Smith}{2008}]{Smith2008a}
Smith, R.~W. (2008).
\newblock {Beyond the Galaxy: The Development of Extragalactic Astronomy
  1885-1965, Part 1}.
\newblock {\em Journal for the History of Astronomy}, 39:91--119.

\bibitem[\protect\astroncite{Stoops}{1958}]{Stoops1958}
Stoops, R., editor (1958).
\newblock {\em {La Structure Et l'{\'{E}}volution de l'Univers: Rapports Et
  Discussions}}.
\newblock Conseil de physique, Instituts Solvay, Bruxelles.

\bibitem[\protect\astroncite{Thorne}{1994}]{Thorne1994}
Thorne, K.~S. (1994).
\newblock {\em {Black holes and time warps: Einstein's outrageous legacy}}.
\newblock W.W. Norton, London.

\bibitem[\protect\astroncite{van Dongen}{2010}]{Dongen2010}
van Dongen, J. (2010).
\newblock {\em {Einstein's Unification}}.
\newblock Cambridge University Press, Cambridge.

\bibitem[\protect\astroncite{Vorontsov-Vel'Yaminov and
  Arkhipova}{1962}]{Vorontsov-VelYaminov1962}
Vorontsov-Vel'Yaminov, B.~A. and Arkhipova, V.~P. (1962).
\newblock {\em {Morphological catalogue of galaxies. Part 1.}}

\bibitem[\protect\astroncite{Weart}{1978}]{Weart1978}
Weart, S. (1978).
\newblock {\em {Interview of Dennis Sciama on 1978 April 14}}.
\newblock Niels Bohr Library {\&} Archives, American Institute of Physics,
  College Park, MD, USA.

\bibitem[\protect\astroncite{Weinberg}{1972}]{Weinberg1972}
Weinberg, S. (1972).
\newblock {\em {Gravitation and Cosmology: Principles and Applications of the
  General Theory of Relativity}}, volume~41.
\newblock John Wiley {\&} Sons, Inc., New York City, NY.

\bibitem[\protect\astroncite{Wheeler}{1957}]{Wheeler1957}
Wheeler, J.~A. (1957).
\newblock {The Present Position of Classical Relativity Theory and Some of its
  Problems}.
\newblock In Rickles, D. and DeWitt-Morette, C., editors, {\em The Role of
  Gravitation in Physics: Report from the 1957 Chapel Hill Conference}, pages
  43--49. Edition Open Sources.

\bibitem[\protect\astroncite{Wheeler}{1958}]{Wheeler1958}
Wheeler, J.~A. (1958).
\newblock {Meaning of the Term, ``Einstein's Theory''}.
\newblock In Stoops, R., editor, {\em La structure et l'{\'{e}}volution de
  l'univers}, pages 98--99. Institut International de Physique Solvay,
  Bruxelles.

\bibitem[\protect\astroncite{Wheeler}{1962}]{Wheeler1962}
Wheeler, J.~A. (1962).
\newblock {The Universe in the Light of General Relativity}.
\newblock {\em Monist}, 47(1):40--76.

\bibitem[\protect\astroncite{Wheeler}{1974}]{Wheeler1974}
Wheeler, J.~A. (1974).
\newblock {The Universe as Home for Man}.
\newblock {\em American Scientist}, 62(6).

\bibitem[\protect\astroncite{Will}{1986}]{Will1986}
Will, C.~M. (1986).
\newblock {\em {Was Einstein right? Putting general relativity to the test.}}
\newblock Basic Books.

\bibitem[\protect\astroncite{Will}{1989}]{Will1989}
Will, C.~M. (1989).
\newblock {The Renaissance of General Relativity}.
\newblock In Davies, P., editor, {\em The New Physics}, pages 7--33. Cambridge
  University Press, Cambridge.

\bibitem[\protect\astroncite{Zwicky}{1933}]{Zwicky1933}
Zwicky, F. (1933).
\newblock {Die Rotverschiebung von extragalaktischen Nebeln}.
\newblock {\em Helvetica Physica Acta}, 6:110--127.

\bibitem[\protect\astroncite{Zwicky et~al.}{1961}]{Zwicky1961}
Zwicky, F., Herzog, E., Wild, P., Karpowicz, M., and Kowal, C.~T. (1961).
\newblock {\em {Catalogue of galaxies and of clusters of galaxies, Vol. I}}.
\newblock California Institute of Technology (CIT), Pasadena.

\end{thebibliography}

\end{document}